\documentclass[12pt]{iopart}
\usepackage{graphicx}
\begin{document}
\title {Dissipative diamagnetism with anomalous coupling and third law.}
\vskip 0.5cm \author{Malay Bandyopadhyay$^1$ and Sushanta Dattagupta$^2$}
\vskip 0.5cm
\address{$^1$Department of Theoretical Physics, Tata Institute of Fundamental Research, Homi Bhabha Road, Colaba, Mumbai-400005, India.\\
$^2$Indian Institute of Science Education \& Research, HC-VII, Salt Lake, Kolkata-700106, India.}
\ead{malay@theory.tifr.res.in}
\begin{abstract}
\vskip 0.5cm
In this work, low temperature thermodynamic behaviour in the context of dissipative diamagnetism with anomalous coupling is analyzed. We find that finite dissipation substitutes the zero-coupling result of exponential decay of entropy by a power law behaviour at low temperature. For Ohmic bath, entropy vanishes linearly with temperature, $T$, in conformity with Nernst's theorem. It is also shown that entropy decays faster in the presence of anomalous coupling than that of the usual coordinate-coordinate coupling. It is observed that velocity-velocity coupling is the most advantageous coupling scheme to ensure the third law of thermodynamics. It is also revealed that different  thermodynamic functions are independent of magnetic field at very low temperature for various coupling schemes discussed in this work. 
\end{abstract}
\maketitle
\section{Introduction}
The third law of thermodynamics is an axiom of nature regarding entropy and the impossibility of reaching absolute zero of temperature. The third law was developed by Walther Nernst and is thus sometimes referred to as Nernst's theorem or Nernst's postulate \cite{a,b}. According to Max Planck, the entropy per particle of an N-body system, $s_0= S/N$, approaches to a constant value and is determined only by the degeneracy of the ground state, $g$, \cite{c}. Thus, the constant value of entropy is given by $S(T=0)=k_B\ln g$, with $k_B$ being the Boltzmann constant. Therefore, the typical value of entropy in the thermodynamic limit ($N\rightarrow \infty$), $s_0=S(T=0)/N$, goes to zero as long as the degeneracy does not grow with $N$ faster than exponentially \cite{d}. This further implies that thermodynamic functions such as entropy, specific heat, the isobaric co-efficient of expansion, the isochoric coefficient of tension etc. all approach zero as $T\rightarrow 0$ \cite{e}.\\
\indent
But, there are certain simple model systems which do not obey the third law. A well known exception is the case of non interacting independent particles each accoutered with spin $I$, yields the limiting entropy per particle, $s_0=k_B\ln (2I+1)$\cite{hanggi}. Another well known example is that of the classical ideal gas for which entropy per particle, $s_0=c_V\ln T+k_B\ln (V/N)+\sigma$, where $V$ is the volume, $c_V$ is the specific heat per particle, and $\sigma$ denotes entropy constant. It clearly shows that the entropy diverges logarithmically with temperature as it goes to zero \cite{hanggi}. Now, proper accounting of degeneracy factor in the form of Fermi-Dirac or Bose-Einstein statistics is able to restore the third law for the above mentioned cases. If we now turn to the case of a free quantum particle for which the specific heat remains constant at $k_b/2$, clearly violates the third law. On the other hand, Einstein oscillator shows an exponential suppression of the specific heat as $T\rightarrow 0$ \cite{ein}. However, Hanggi and Ingold have shown that the low temperature behaviour for the above mentioned two cases changes qualitatively  when the system is strongly coupled to a quantum mechanical heat bath \cite{hanggi, hanggi1}. This finite dissipation ensures linear decay of entropy with temperature for the Einstein oscillator as well as for the free quantum particle as $T\rightarrow 0$. This observation enables Hanngi and Ingold to arrive at an interesting conclusion that quantum statistics is just the first step to ensure third law and a more crucial step to satisfy third law is to make the system ``open" i.e. the system needs to be strongly coupled to a quantum heat bath. This conclusion is not only interesting in the academic perspective but is relevant for experiments in nanosystems which are strongly influenced by the environment for their smallness and large surface to volume ratio \cite{mazenko,datta,imry,chakravarty}.\\
\indent
Diamagnetism is an intrinsically quantum mechanical property, the treatment for which was first discussed by Landau, considering a collection of electrons in a box in the presence of an external constant magnetic field \cite{landau}. In this context the essential role of quantum mechanics as well as the role of boundary is discussed in details by van Vleck and Peierls \cite{van,peierls}. The boundary effects can be recovered by using a parabolic potential characterized by a frequency $\omega_0$, a trick introduced by Darwin \cite{darwin}. The essential role of boundary electrons was further clarified in the context of dissipative diamagnetism by several authors \cite{datta1,datta2,malay}. As we are discussing third law of thermodynamics in the context of dissipative diamagnetism, we discuss separately the $\omega_0=0$ and $\omega_0\neq 0$ cases. Since the parabolic potential which considered here is physically realizable in a quantum dot or in a quantum well nanostructure, the results for $\omega_0\neq 0$ are of independent interest \cite{merkt}.\\  
\indent
Here, we investigate low temperature thermodynamic properties in the context of dissipative diamagnetism with different anomalous coupling schemes. For that purpose we consider a charged quantum particle in the presence of an external constant magnetic field when it is in contact with a dissipative quantum heat bath. This kind of analysis is related with the dissipative quantum mechanics, a subject that has seen a great attention through the work of Leggett and others \cite{legg1,legg2,legg3}. There are several approaches for the treatment of dissipative quantum systems. The most conventional approach is system plus reservoir point of view i.e. the system of interest is coupled linearly with the environment which is represented by a collection of harmonic oscillators \cite{li1,li2}. Usually, one is interested in the dissipative subsystem and the reservoir variables are eliminated by projection operator or tracing procedure \cite{sch}. As a result of that, the reservoir enters only through few parameters. The results obtained from these kind of dissipative quantum systems are of great interest due to the recent widespread interest on the critical role of environmental effects in mesoscopic systems \cite{mazenko,datta,imry,chakravarty}, in fundamental quantum physics, and in quantum information \cite{bennett,zeilinger,giulini,myatt}. All these recent developments in the subject of quantum thermodynamics \cite{capek,shicka,sheehan,allah} and widespread interest on the low temperature physics of small quantum systems has raised up the question : Does the third law of thermodynamics hold in the quantum regime? How quantum dissipation can play an important role in thermodynamic theory? Several authors have tried to settle all these issues. Ford and O'Connell discussed about the third law of thermodynamics in connection with a quantum harmonic oscillator \cite{ford}. Recently, P. Hanggi and G. L. Ingold have shown that finite dissipation actually helps to ensure the third law of thermodynamics \cite{hanggi,hanggi1}. Further investigations has been made by W. C. Yang and B. J. Dong on the influence of various coupling forms for a harmonic oscillator \cite{bao}. The third law has also been validated for a charged magneto-oscillator \cite{malay1}. Similar kind of analysis in the context of dissipative diamagnetism can be found in \cite{jishad}. In this work, we have extended all the above mentioned studies by considering dissipative diamagnetism with different anomalous coupling which not only demonstrate the environmental effect in nanostructure but also illustrate the essential role of boundary. \\
\indent
With this preceding background, we organize the rest of the paper as follows. In the next section, we introduce the model system and different coupling schemes. In Sec. III, we discuss coordinate-coordinate coupling scheme. In this connection, without dissipation and free quantum particle cases are also analyzed. In addition, explicit results of low temperature thermodynamical quantities are derived analytically for the Ohmic model, Lorentzian power spectrum model, exponentially correlated model, and arbitrary heat bath model. Coordinate (velocity)- velocity (coordinate) scheme is examined in details in Sec. IV. In this connection, the radiation heat bath case is also analyzed. Section V deals with the velocity-velocity coupling scheme. For all the above mentioned coupling schemes, the cases for $\omega_0=0$ and $\omega_0\neq 0$ are analyzed separately. Finally, we conclude this paper in Sec. VI.
\vskip 0.5cm
{\section{Model System}}
The starting point of this section is the generalized Caldeira-Legget system-plus-reservoir Hamiltonian for a charged particle of mass `$m$' and charge `$e$' in a magnetic field $\vec{B}$ in the operator form \cite{l,m} :
\begin{eqnarray}
\hat{H}=\frac{\Big(\hat{p}-e\vec{A}/c\Big)^2}{2m}+V(\hat{r})+\sum_{j=1}^N\Big\lbrack\frac{1}{2m_j}\Big(\hat{p}_j^2+m_j^2\omega_j^2\hat{q}_j^2\Big)+g(\hat{r},\hat{p},\hat{q}_j,\hat{p}_j)\Big\rbrack,
\end{eqnarray}
where $\lbrace\hat{r},\hat{p}\rbrace$ and $\lbrace\hat{q}_j,\hat{p}_j\rbrace$ are the sets of co-ordinate and momentum operators of system and bath oscillators. They follow the following commutation relations
\begin{eqnarray}
\lbrack \hat{r}_{\alpha},\hat{p}_{\beta}\rbrack = i\hbar\delta_{\alpha\beta},
\lbrack\hat{q}_{i\alpha},\hat{p}_{j\beta}\rbrack=i\hbar\delta_{ij}\delta_{\alpha\beta},
\end{eqnarray}
where $\alpha$, $\beta$ denote components of the above mentioned operators along $x$, $y$ direction respectively. Equation (1) includes four types of bilinear coupling between the system and the environmental degrees of freedom. For the usual coordinate-coordinate coupling \cite{l},
\begin{equation}
g=-c_j\hat{r}\hat{q}_j+\frac{c_j^2\hat{r}^2}{2m_j\omega_j^2},
\end{equation}
for the system coordinate and environmental velocity coupling \cite{m},
\begin{equation}
g=-d_{1,j}\frac{\hat{r}\hat{p}_j}{m_j}+\frac{d_{1,j}^2\hat{r}^2}{2m_j},
\end{equation}
or for the system velocity and environmental coordinate coupling \cite{n},
\begin{equation}
g=-d_{2,j}\frac{\hat{p}\hat{q}_j}{m}+\frac{d_{2,j}^2\hat{q}_j^2}{2m},
\end{equation} 
and finally for system velocity and environmental velocity coupling \cite{o},
\begin{equation}
g=-e_{j}\frac{\hat{p}\hat{p}_j}{mm_j}+\frac{e_{j}^2\hat{p}_j^2}{2mm_j^2}.
\end{equation}
The additional terms appearing in the coupling are in order to compensate coupling induced potential and mass renormalization.\\
Now, we are interested in investigating low temperature thermodynamic behaviour of the model system described above. For this purpose, we need to calculate the free energy of the system exactly. This is a non-trivial quantity to calculate and the details of its derivation can be found in \cite{n,ford1,ford2}. The free energy of the charged magneto-oscillator is given by \cite{p,r,s},
\begin{eqnarray}
F =\frac{1}{\pi}\int_0^{\infty}d\omega f(\omega,T)\Im\Big\lbrack\frac{d}{d\omega}\ln\Big(\det\alpha(\omega+i0^+)\Big)\Big\rbrack,
\end{eqnarray}
where $\alpha(\omega)$ denotes the generalized susceptibility of the model system and $f(\omega,T)$ is the free energy of a single oscillator of frequency $\omega$ and is given by 
\begin{equation}
f(\omega,T)=k_BT\log\Big\lbrack 1 - \exp(-\frac{\hbar\omega}{k_BT})\Big\rbrack,
\end{equation}
where we have ignored the zero-point contribution. This formula is remarkable in the sense that it expresses oscillator free energy exactly in terms of single integration over the oscillator susceptibility alone.  Now, we can rewrite Eq. (7) as follows \cite{r,s}:
\begin{equation}
F(T,B)=F(T,0)+\Delta F(T,B),
\end{equation}
where 
\begin{equation}
F(T,0)=\frac{3}{\pi}\int_0^{\infty}d\omega f(\omega, T)I_1
\end{equation}
is the free energy of the oscillator in the absence of the magnetic field, $I_1=\Im\Big\lbrack\frac{d}{d\omega}\ln\alpha^{(0)}(\omega)\Big\rbrack$, $\alpha^{(0)}(\omega)$ is the scalar susceptibility in the absence of a magnetic field and the correction due to the magnetic field is given by
\begin{equation}
\Delta F(T,B)=-\frac{1}{\pi}\int_0^{\infty}d\omega f(\omega, T)I_2,
\end{equation}
where $I_2=\Im\Big\lbrace\frac{d}{d\omega}\ln\Big\lbrack 1-\Big(\frac{eB\omega\alpha^{(0)}}{c}\Big)^2\Big\rbrack\Big\rbrace$. The function $f(\omega,T)$ vanishes exponentially for $\omega>>\frac{k_BT}{\hbar}$ and hence all the integrand in Eq. (10) and in Eq. (11) are confined to low frequencies. Thus, one can easily obtain an explicit results by expanding the factor multiplying $f(\omega, T)$ in powers of $\omega$.\\
The scalar susceptibility for a harmonic oscillator in the absence of a magnetic field is given by \cite{r,s}
\begin{equation}
\alpha^{(0)}(\omega)=\frac{1}{m(\omega_0^2-\omega^2)-i\omega\tilde{\gamma}_{\mu}(\omega)},
\end{equation}
where
\begin{equation}
\tilde{\gamma}_{\mu}(\omega)=\int_0^tdt^{\prime}\gamma_{\mu}(t^{\prime})e^{i\omega t^{\prime}}.
\end{equation}
Here $\mu=1,2,3,4$ and denotes the subscript for four different coupling schemes. The memory kernel is given by
\begin{equation}
\gamma_{\mu}(t)=\frac{2}{m\pi}\int_0^{\infty}d\omega \frac{J_{\mu}(\omega)}{\omega}\cos(\omega t).
\end{equation} 
In equation (14), $J_{\mu}(\omega)$ denotes the spectral density function of the heat bath oscillators for different type of coupling schemes and is given as follows:
\begin{eqnarray}
J_1(\omega)=J_{c-c}(\omega)=\pi\sum_{j=1}^N\frac{c_j^2}{2m_j\omega_j}\delta(\omega-\omega_j),\\
J_2(\omega)=J_{c-v}(\omega)=\pi\sum_{j=1}^N\frac{d_{1,j}^2}{2m_j}\omega_j\delta(\omega-\omega_j),\\
J_3(\omega)=J_{v-c}(\omega)=\pi\sum_{j=1}^N\frac{d_{2,j}^2}{2m_j}\omega_j\delta(\omega-\omega_j),\\
J_4(\omega)=J_{v-v}(\omega)=\pi\sum_{j=1}^N\frac{e_j^2}{2m_j}\omega_j^3\delta(\omega-\omega_j).
\end{eqnarray}
\begin{figure}[h]
\begin{center}
{\rotatebox{270}{\resizebox{7cm}{12cm}{\includegraphics{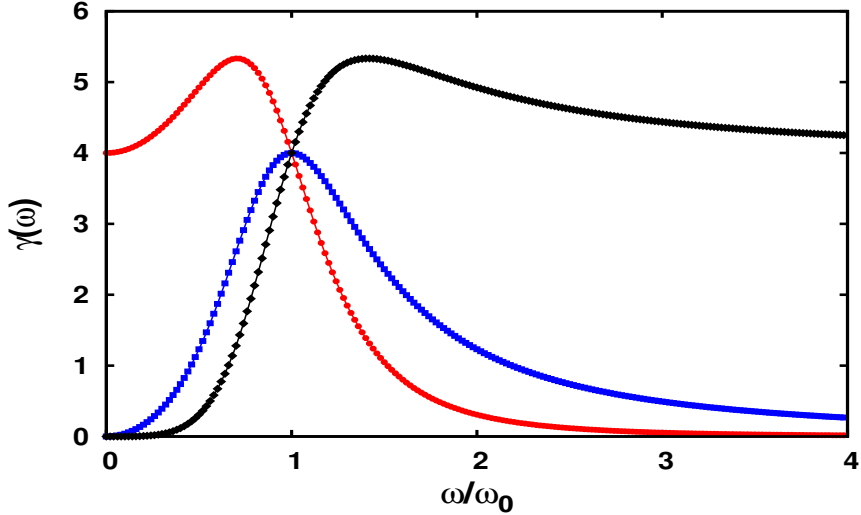}}}}
\caption{(color online) Plot of $\tilde{\gamma}(\omega)$ versus dimensionless frequency $\frac{\omega}{\omega_0}$ for the coordinate-coordinate coupling scheme (red filled circle), coordinate-velocity or velocity-coordinate coupling scheme (blue filled square) and for the velocity-velocity coupling scheme (black filled diamond). To plot this figure, we use $\frac{\Gamma}{\Omega}=1.0$, $m=1.0$, and $\gamma=1.0$.}
\end{center}
\end{figure} 
To show the distinct behaviour of different kind of coupling schemes, we plot the power spectra of the memory friction function in Fig.1 for the coordinate-coordinate (c-c), coordinate-velocity (c-v), and velocity-velocity (v-v) coupling schemes. \\
We can calculate $I_1$ and $I_2$ explicitly. The expressions are given as follows :
\begin{equation}
I_1=\frac{m\tilde{\gamma}_{\mu}(\omega)(\omega_0^2+\omega^2)+m\omega\tilde{\gamma}^{\prime}_{\mu}(\omega)(\omega_0^2-\omega^2)}{\lbrack m^2(\omega_0^2-\omega^2)^2+\omega^2\tilde{\gamma}_{\mu}^2(\omega)\rbrack},
\end{equation}
and 
\begin{eqnarray}
\hskip-0.8cm
&&I_2=2\frac{m\tilde{\gamma}_{\mu}(\omega)(\omega_0^2+\omega^2)+m\omega\tilde{\gamma}^{\prime}_{\mu}(\omega)(\omega_0^2-\omega^2)}{\lbrack m^2(\omega_0^2-\omega^2)^2+\omega^2\tilde{\gamma}_{\mu}^2(\omega)\rbrack} \nonumber \\
\hskip-0.8cm
&&-\frac{m\tilde{\gamma}_{\mu}(\omega)(\omega_0^2+\omega^2)+m\omega\tilde{\gamma}_{\mu}^{\prime}(\omega)(\omega_0^2-\omega^2+\omega\omega_c)}{\lbrack m^2(\omega_0^2-\omega^2+\omega\omega_c)^2+\omega^2\tilde{\gamma}_{\mu}^2(\omega)\rbrack} \nonumber \\
\hskip-0.8cm
&&-\frac{m\tilde{\gamma}_{\mu}(\omega)(\omega_0^2+\omega^2)+m\omega\tilde{\gamma}_{\mu}^{\prime}(\omega)(\omega_0^2-\omega^2-\omega\omega_c)}{\lbrack m^2(\omega_0^2-\omega^2-\omega\omega_c)^2+\omega^2\tilde{\gamma}_{\mu}^2(\omega)\rbrack},
\end{eqnarray}
where $\omega_c=\frac{eB}{mc}$ is the cyclotron frequency. Thus, our main task is to find free energy $F$  at low temperature. Then, one can easily derive other thermodynamic functions at low temperature. For example, entropy is defined as 
\begin{equation}
S=-\frac{\partial F}{\partial T}.
\end{equation}
We have now all the essential ingredients to calculate thermodynamic functions. \\
\section{Coordinate-coordinate coupling scheme}
First, let us consider the usual case of the system's coordinate coupled with the coordinates of the heat bath. This kind of coupling can be realized experimentally in the case of a RLC circuit driven by a Gaussian white noise. In this connection, we analyze the free particle and without dissipation case. In addition, we examine the decay behaviour of entropy with temperature for Ohmic, exponentially correlated, and arbitrary heat bath models. The results for $\omega_0=0$ and $\omega \neq 0$ are separately discussed for all the above mentioned cases. \\
\subsection{Without Dissipation}
The limit of without dissipation can easily be obtained by taking $\tilde{\gamma}_{\mu}(\omega)=0$. Thus, 
\begin{equation}
\alpha^{(0)}(\omega)=-\frac{1}{m(\omega^2-\omega_0^2)},
\end{equation}
and
\begin{equation}
\Big\lbrack 1-\Big(\frac{eB\omega}{c}\Big)^2\lbrack\alpha^{(0)}(\omega)\rbrack^2\Big\rbrack=\frac{\Big\lbrack (\omega^2-\omega_0^2)^2-(\omega\omega_c)^2\Big\rbrack}{(\omega^2-\omega_0^2)^2},
\end{equation}
where $\omega_c=\frac{eB}{mc}$ is the cyclotron frequency. For this case, 
\begin{equation}
I_1=\Im\Big\lbrace\frac{d}{d\omega}\ln\alpha^{(0)}(\omega)\Big\rbrace =\pi\Big\lbrack\delta(\omega-\omega_0)+\delta(\omega+\omega_0)\Big\rbrack,
\end{equation}
where we have used the identity
\begin{equation}
\frac{1}{\omega-\omega_j+i0^+}=P\Big\lbrack\frac{1}{\omega-\omega_j}\Big\rbrack-i\pi\delta(\omega-\omega_j).
\end{equation}
Thus,
\begin{equation}
F(T,0)=3f(\omega_0,T)
\end{equation} 
Similarly, one can show that
\begin{equation}
\Delta F(T,B)=f(\omega_1,T)+f(\omega_2,T)-2f(\omega_0,T),
\end{equation}
where $\omega_{1,2}=\pm\frac{\omega_c}{2}+\lbrack \omega_0^2+(\frac{\omega_c}{2})^2\rbrack^{\frac{1}{2}}$. Hence
\begin{equation}
F(T,B)=f(\omega_0,T)+f(\omega_1,T)+f(\omega_2,T).
\end{equation}
At low temperature free energy becomes
\begin{equation}
F(T,B)=-k_BT\Big(e^{-\frac{\hbar\omega_0}{k_BT}}+e^{-\frac{\hbar\omega_1}{k_BT}}+e^{-\frac{\hbar\omega_2}{k_BT}}\Big).
\end{equation}
Finally, entropy of the system is given by
\begin{eqnarray}
\hskip -1.0cm
S(T,B)=k_B\Big\lbrack(1+\frac{\hbar\omega_0}{k_BT})e^{-\frac{\hbar\omega_0}{k_BT}}+(1+\frac{\hbar\omega_1}{k_BT})e^{-\frac{\hbar\omega_1}{k_BT}}(1+\frac{\hbar\omega_2}{k_BT})e^{-\frac{\hbar\omega_2}{k_BT}}\Big\rbrack.
\end{eqnarray}
Thus, it can be concluded that entropy, $S(T)$, vanishes exponentially when $T\rightarrow 0$ for the without dissipation case.\\
\subsection{Ohmic heat bath}
For pure Ohmic heat bath, one can take $\tilde{\gamma}_1(\omega)=m\gamma$, where $\gamma$ is friction constant. Thus, the response function in the absence of magnetic field becomes
\begin{equation}
\alpha^{(0)}(\omega)=\Big\lbrack m(\omega_0^2-\omega^2)-im\omega\gamma\Big\rbrack^{-1}.
\end{equation} 
Thus,
\begin{eqnarray*}
I_1=\frac{\gamma(\omega^2+\omega_0^2)}{(\omega^2-\omega_0^2)^2+\gamma^2\omega^2}
\stackrel{\omega\rightarrow 0}{\simeq}\frac{\gamma}{\omega_0^2}.
\end{eqnarray*}
Similarly,
\begin{eqnarray*}
\hskip-1.5cm
-I_2&=&\frac{\gamma(\omega^2+\omega_0^2)}{(\omega^2-\omega_0^2+\omega\omega_c)^2+\gamma^2\omega^2}+ \frac{\gamma(\omega^2+\omega_0^2)}{(\omega^2-\omega_0^2-\omega\omega_c)^2+\gamma^2\omega^2}-2\frac{\gamma(\omega^2+\omega_0^2)}{(\omega^2-\omega_0^2)^2+\gamma^2\omega^2} \\
\hskip-1.5cm
&&\stackrel{\omega\rightarrow 0}{\simeq}\frac{\gamma}{\omega_0^2}+\frac{\gamma}{\omega_0^2}-2\frac{\gamma}{\omega_0^2} = 0.
\end{eqnarray*}
Hence, free energy of the system at low temperature can be written as
\begin{equation}
F(T)=\frac{3k_BT\gamma}{\pi\omega_0^2}\int_0^{\infty}d\omega\ln\Big\lbrack 1-\exp\big(-\frac{\hbar\omega}{k_BT}\Big)\Big\rbrack.
\end{equation}
The following integral is relevant for our calculation throughout this paper: 
\begin{equation}
\int_0^{\infty}dy y^{\nu}\log(1-e^{-y})=-\Gamma(\nu+1)\zeta(\nu+2),
\end{equation}
where $\Gamma$ is gamma function and $\zeta$ is Riemann's zeta function.
Using this integral, one can show
\begin{equation}
F(T)=-\frac{\pi}{2}\hbar\gamma\Big(\frac{k_BT}{\hbar\omega_0}\Big)^2.
\end{equation}
Hence, entropy is given by
\begin{equation}
S(T)=\pi\hbar\gamma\frac{k_B^2T}{(\hbar\omega_0)^2}.
\end{equation}
As $T\rightarrow 0$, $S(T)$ vanishes linearly with $T$ which perfectly matches with third law of thermodynamics. It shows the usual Ohmic friction behaviour of linear decay. Also, one can notice that decay slope is directly proportional to $\gamma$.\\
Now, one can consider the case for $\omega_0=0$ for the same charged particle in an external constant magnetic field in contact with a Ohmic bath. For this case,
\begin{equation}
I_2(\omega)=-2\frac{\gamma\omega^2}{(\omega^2-\omega\omega_c)^2+\gamma^2\omega^2}\stackrel{\omega\rightarrow 0}{\simeq}\frac{-2\gamma}{\gamma^2+\omega_c^2},
\end{equation}
and $I_1=0$. Thus, the free energy becomes
\begin{eqnarray}
F(T)&=&\frac{2k_BT\gamma}{\pi(\omega_c^2+\gamma^2)}\int_0^{\infty}d\omega\ln\Big\lbrack 1-\exp\big(-\frac{\hbar\omega}{k_BT}\Big)\Big\rbrack \nonumber \\.
&=&-\frac{\pi}{3\hbar}\frac{\gamma}{\gamma^2+\omega_c^2}(k_BT)^2
\end{eqnarray}
The decay behaviour of entropy follows
\begin{equation}
S(T)=\frac{2\pi}{3\hbar}\frac{\gamma}{\gamma^2+\omega_c^2}k_B^2T.
\end{equation}
\subsection{free particle}
Next, we consider a free quantum Brownian particle in contact with a Ohmic heat bath for which 
\begin{equation}
\alpha^{(0)}(\omega)=\lbrack -m\omega^2-im\omega \gamma\rbrack^{-1}.
\end{equation}
Thus, we have
\begin{equation*}
I_1= \frac{m^2\gamma\omega^2}{(m^2\omega^4+m^2\omega^2\gamma^2}\stackrel{\omega\rightarrow 0}{\simeq}\frac{1}{\gamma},
\end{equation*}
and $I_2=0$. Free energy for the free Brownian particle is given by
\begin{equation}
F(T)=-\frac{\pi}{2}\hbar\gamma\Big(\frac{k_BT}{\hbar\gamma}\Big)^2,
\end{equation}
and the decay behaviour of entropy becomes \cite{jishad}
\begin{equation}
S(T)=\frac{\pi k_B^2T}{\hbar\gamma}.
\end{equation}
Unlike equation (35), the slope of the entropy for free particle is inversely proportional to $\gamma$.
\subsection{Lorentzian power spectrum} 
Now, we consider that the environmental oscillators have a power spectrum with a narrow Lorentzian peak centered at a finite frequency not at zero. Thus, the Fourier transform of the memory function is
\begin{equation}
\tilde{\gamma}_1(\omega)=\frac{m\gamma\Omega^4}{\Gamma^2\omega^2+(\Omega^2-\omega^2)^2},
\end{equation}
where $\gamma$ denotes the Markovian friction strength of the system, $\Gamma$ and $\Omega$ are the damping and frequency parameters of the harmonic noise \cite{t}. Now,
\begin{eqnarray*}
&&I_1=\frac{m^2\gamma\Omega^4DA-m^2\gamma\Omega^4\omega A^{\prime}C}{m^2C^2A^2+4m^2\gamma^2\Omega^8\omega^2}\\
&&\stackrel{\omega\rightarrow 0}{\simeq}\frac{\gamma}{\omega_0^2},
\end{eqnarray*}
where $A=\lbrack\Gamma^2\omega^2+(\Omega^2-\omega^2)^2\Big\rbrack$, $A^{\prime}=\frac{dA}{d\omega}$, $C=(\omega_0^2-\omega^2)$ and $D=(\omega^2+\omega_0^2)$ and
\begin{eqnarray*}
\hskip-0.8cm
-I_2&=&\frac{m^2\gamma\Omega^4DA-m^2\gamma\Omega^4\omega A^{\prime}C_1}{m^2C_1^2A^2+4m^2\gamma^2\Omega^8\omega^2}+\frac{m^2\gamma\Omega^4DA-m^2\gamma\Omega^4\omega A^{\prime}C_2}{m^2C_2^2A^2+4m^2\gamma^2\Omega^8\omega^2}\\
\hskip -0.8cm
&&-2\frac{m^2\gamma\Omega^4DA-m^2\gamma\Omega^4\omega A^{\prime}C}{m^2C^2A^2+4m^2\gamma^2\Omega^8\omega^2}\\
\hskip-0.8cm
&&\stackrel{\omega\rightarrow 0}{\simeq}\frac{\gamma}{\omega_0^2}+\frac{\gamma}{\omega_0^2}-2\frac{\gamma}{\omega_0^2}= 0,
\end{eqnarray*}
where $C_1=(\omega^2-\omega_0^2+\omega\omega_c)$ and $C_2=(\omega^2-\omega_0^2-\omega\omega_c)$. Thus free energy of our model system for the Ohmic heat bath with Lorentzian peak power spectrum is given by
\begin{equation}
F(T)= -\frac{\pi}{2}\hbar\gamma\Big(\frac{k_BT}{\hbar\omega_0}\Big)^2.
\end{equation}
Hence, the decay behaviour of entropy is again the same as of Eq. (35).\\
Let us discuss the case of $\omega_0=0$ for the same kind of Lorentzian spectrum. For this case $I_2\stackrel{\omega\rightarrow 0}{\simeq}-\frac{2\gamma(1-\gamma/\Omega)}{\gamma^2+\omega_c^2}$ and $I_1=0$. Thus, free energy of this charged magneto-oscillator with $\omega_0=0$ and in contact to a heat bath with Lorentzian spectrum is given by
\begin{equation}
F(T)= -\frac{\pi\gamma(1-\gamma/\Omega)}{3\hbar(\gamma^2+\omega_c^2)}(k_BT)^2,
\end{equation}
and entropy is given by
\begin{equation}
S(T)=\frac{2\pi\gamma(1-\gamma/\Omega)}{3\hbar(\gamma^2+\omega_c^2)}k_B^2 T.
\end{equation}
\subsection{Exponentially correlated heat bath}
In this subsection, we consider exponentially correlated heat bath whose memory friction is given by
\begin{equation}
\gamma_1(t)=\frac{m\gamma}{\tau_c}e^{-\frac{|t|}{\tau_c}}.
\end{equation}
The Fourier transform of memory friction gives us
\begin{equation}
\tilde{\gamma}_1(\omega)=\frac{m\gamma}{1+\omega^2\tau_c^2}.
\end{equation}
Now, the required expressions for $I_1$ and $I_2$ are as follows
\begin{eqnarray*}
&&I_1=\frac{m^2\gamma(1+\omega^2\tau_c^2)D-2m^2\omega^2\gamma\tau_c^2C}{m^2C^2(1+\omega^2\tau_c^2)^2+m^2\gamma^2\omega^2}\\
&&\stackrel{\omega\rightarrow 0}{\simeq}\frac{\gamma}{\omega_0^2}.
\end{eqnarray*}
Similarly,
\begin{eqnarray*}
\hskip-1.5cm
-I_2&=&\frac{m^2\gamma(1+\omega^2\tau_c^2)D-2m^2\omega^2\gamma\tau_c^2C_1}{m^2C_1^2(1+\omega^2\tau_c^2)^2+m^2\gamma^2\omega^2}+\frac{m^2\gamma(1+\omega^2\tau_c^2)D-2m^2\omega^2\gamma\tau_c^2C_2}{m^2C_2^2(1+\omega^2\tau_c^2)^2+m^2\gamma^2\omega^2}\\
\hskip-1.5cm
&&-2\frac{m^2\gamma(1+\omega^2\tau_c^2)D-2m^2\omega^2\gamma\tau_c^2C}{m^2C^2(1+\omega^2\tau_c^2)^2+m^2\gamma^2\omega^2}\\
\hskip-1.5cm
&&\stackrel{\omega\rightarrow 0}{\simeq}\frac{\gamma}{\omega_0^2}+\frac{\gamma}{\omega_0^2}-2\frac{\gamma}{\omega_0^2} = 0.
\end{eqnarray*}
Thus, free energy of this model system with exponentially correlated heat bath is given by
\begin{equation}
F(T)= -\frac{\pi}{2}\hbar\gamma\Big(\frac{k_BT}{\hbar\omega_0}\Big)^2.
\end{equation}
Hence, the decay behaviour of entropy with temperature is again linear which is same as that of Ohmic model. Now, we discuss the case without the confining potential, $\omega_0=0$. For the charged particle in a magnetic field in contact with a exponentially correlated heat bath :
\begin{equation}
I_2(\omega)\stackrel{\omega\rightarrow 0}{\simeq}-\frac{2\gamma(1-\gamma\tau_c)}{(\gamma^2+\omega_c^2)},
\end{equation}
and $I_1=0$. Thus, the free energy becomes
\begin{equation}
F(T)= -\frac{\pi\gamma(1-\gamma\tau_c)}{3\hbar(\gamma^2+\omega_c^2)}(k_BT)^2,
\end{equation}
and entropy is given by
\begin{equation}
S(T)=\frac{2\pi\gamma(1-\gamma\tau_c)}{3\hbar(\gamma^2+\omega_c^2)}k_B^2 T.
\end{equation}
This result matches with that of Jishad et al \cite{jishad}.
\subsection{Arbitrary heat bath}
The heat bath is characterized by the memory friction function $\tilde{\gamma}(z)$. According to Ford {\it et al} \cite{ford}, it should be positive and must be analytic in the upper half plane and must satisfy the reality condition
\begin{equation}
\tilde{\gamma}_1(-\omega+i0^+)=\tilde{\gamma}(\omega+i0^+).
\end{equation}
Now, according to Ford {\it et al} \cite{ford} the memory function must be in the neighbourhood of origin as follows :
\begin{equation}
\tilde{\gamma}_1(\omega)\simeq m b^{1-\nu}(-i\omega)^{\nu},
\end{equation}
where $-1<\nu<1$, and $b$ is a positive constant with dimensions of frequency. Thus, the scalar susceptibility of the model system in the absence of the magnetic field is given by
\begin{equation}
\alpha^{(0)}(\omega)=\frac{1}{m(\omega_0^2-\omega^2)+mb^{1-\nu}(-i\omega)^{1+\nu}}.
\end{equation}
Thus, 
\begin{eqnarray*}
I_1&=&\frac{b^{1-\nu}\omega^{\nu}\cos(\frac{\nu\pi}{2})\Big\lbrack(1+\nu)C+2\omega^2\Big\rbrack}{\Big|C+b^{1-\nu}(-i\omega)^{1+\nu}\Big|^2}\\
&&\stackrel{\omega\rightarrow 0}{\simeq}(1+\nu)\cos(\frac{\nu\pi}{2})\frac{b^{1-\nu}}{\omega_0^2}\omega^{\nu},
\end{eqnarray*}
and 
\begin{eqnarray*}
\hskip-1.5cm
-I_2&=&\frac{b^{1-\nu}\omega^{\nu}\cos(\frac{\nu\pi}{2})\Big\lbrack(1+\nu)C_1+2\omega^2\Big\rbrack}{\Big|C_1+b^{1-\nu}(-i\omega)^{1+\nu}\Big|^2}+\frac{b^{1-\nu}\omega^{\nu}\cos(\frac{\nu\pi}{2})\Big\lbrack(1+\nu)C_2+2\omega^2\Big\rbrack}{\Big|C_2+b^{1-\nu}(-i\omega)^{1+\nu}\Big|^2}\\
\hskip-1.5cm
&&-2\frac{b^{1-\nu}\omega^{\nu}\cos(\frac{\nu\pi}{2})\Big\lbrack(1+\nu)C+2\omega^2\Big\rbrack}{\Big|C+b^{1-\nu}(-i\omega)^{1+\nu}\Big|^2}\\
\hskip-1.5cm
&&\stackrel{\omega\rightarrow 0}{\simeq}2(1+\nu)\cos(\frac{\nu\pi}{2})\frac{b^{1-\nu}}{\omega_0^2}\omega^{\nu}-2(1+\nu)\cos(\frac{\nu\pi}{2})\frac{b^{1-\nu}}{\omega_0^2}\omega^{\nu}=0.
\end{eqnarray*}
Hence, free energy of the model system with such arbitrary heat bath is given by
\begin{equation}
F(T)=-3\Gamma(\nu+2)\zeta(\nu+2)\cos\Big(\frac{\nu\pi}{2}\Big)\frac{\hbar b^3}{\pi\omega_0^2}\Big(\frac{k_BT}{\hbar b}\Big)^{2+\nu}.
\end{equation}
Finally, entropy of the system is 
\begin{equation}
S(T)=3\Gamma(\nu+3)\zeta(\nu+2)\cos\Big(\frac{\nu\pi}{2}\Big)\frac{k_Bb^2}{\pi\omega_0^2}\Big(\frac{k_BT}{\hbar b}\Big)^{1+\nu}.
\end{equation}
Since $(\nu+1)$ is always positive, entropy, $S(T)$, vanishes as $T\rightarrow 0$.
\section{Coordinate (Velocity) - Velocity (Coordinate) Coupling}
In this section, we consider the typical case of the first kind where coordinate (velocity) of the system coupled with the velocities (coordinates) of the heat bath. The physical situations for such scheme can be found for a vortex transport in the presence of magnetic field \cite{u} and particle interacting via dipolar coupling \cite{v}. The friction memory function for this kind of coupling scheme is given by
\begin{equation}
\gamma_2(t)=m\gamma\Gamma\exp\Big(-\frac{\Gamma t}{2}\Big)\Big(\cos(\omega_1t)-\frac{\Gamma}{2\omega_1}\sin(\omega_1 t)\Big).
\end{equation}
where $\omega_1^2=\Omega^2-\frac{\Gamma^2}{4}$, $\gamma$ is the Markovian friction strength, $\Gamma$ and $\Omega$ are the damping and frequency parameters of the harmonic noise. The Fourier transform of the memory friction function is given by
\begin{equation}
\tilde{\gamma}_2(\omega)=\frac{2m\gamma\Gamma^2\omega^2}{\Gamma^2\omega^2+(\Omega^2-\omega^2)^2}.
\end{equation}
The expressions of $I_1$ and $I_2$ for this particular scheme are given by 
\begin{eqnarray*}
&&I_1=\frac{2m^2\gamma\Gamma^2\omega^2DA+4m^2\gamma\Gamma^2\omega^2AC-2m^2\gamma\Gamma^2\omega^3A^{\prime}C}{m^2(AC)^2+4m^2\gamma^2\Gamma^4\omega^6}\\
&&\stackrel{\omega\rightarrow 0}{\simeq}\frac{6\gamma\Gamma^2}{\Omega^4\omega_0^2},
\end{eqnarray*}
and
\begin{eqnarray*}
-I_2&=&\frac{2m^2\gamma\Gamma^2\omega^2DA+4m^2\gamma\Gamma^2\omega^2AC_1-2m^2\gamma\Gamma^2\omega^3A^{\prime}C_1}{m^2(AC_1)^2+4m^2\gamma^2\Gamma^4\omega^6}\\
&&+\frac{2m^2\gamma\Gamma^2\omega^2 DA+4m^2\gamma\Gamma^2\omega^2AC_2-2m^2\gamma\Gamma^2\omega^3A^{\prime}C_2}{m^2(AC_2)^2+4m^2\gamma^2\Gamma^4\omega^6}\\
&&-2\frac{2m^2\gamma\Gamma^2\omega^2DA+4m^2\gamma\Gamma^2\omega^2AC-2m^2\gamma\Gamma^2\omega^3A^{\prime}C}{m^2(AC)^2+4m^2\gamma^2\Gamma^4\omega^6}\\
&&\stackrel{\omega\rightarrow 0}{\simeq}\frac{6\gamma\Gamma^2}{\Omega^4\omega_0^2}+\frac{6\gamma\Gamma^2}{\Omega^4\omega_0^2}-12\frac{6\gamma\Gamma^2}{\Omega^4\omega_0^2}=0.
\end{eqnarray*}
\begin{figure}[h]
\begin{center}
{\rotatebox{270}{\resizebox{7cm}{12cm}{\includegraphics{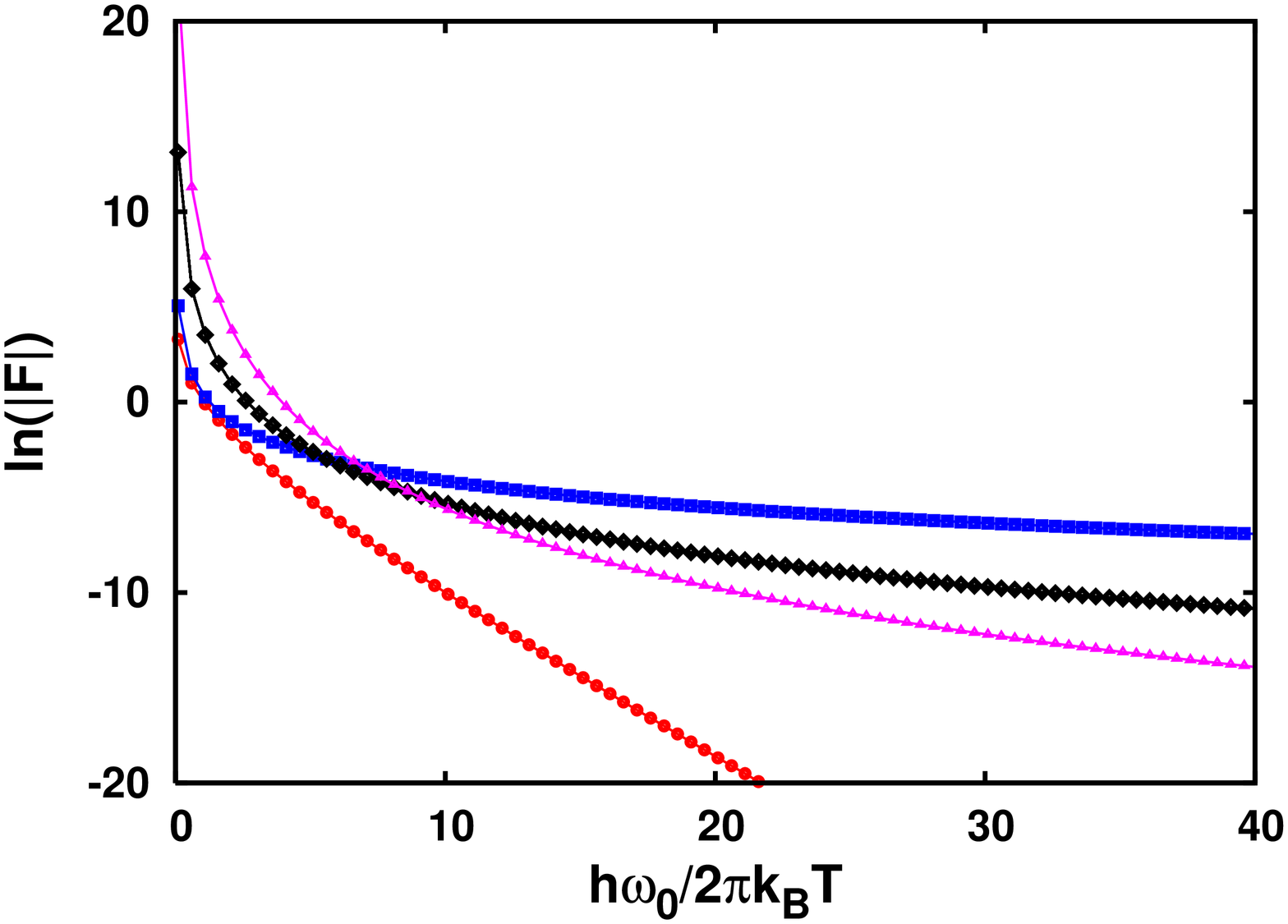}}}}
\caption{(color online) Plot of $\ln(|F|)$ versus dimensionless inverse-temperature $\frac{h\omega_0}{2\pi k_BT}$ for the without dissipation case (red filled circle), for the coordinate-coordinate (c-c) coupling (blue filled square), for the coordinate-velocity (c-v) coupling (balck filled diamond) and for the velocity-velocity (v-v) coupling (pink filled triangle) schemes. To plot this figure, we use $\frac{\hbar\gamma}{\hbar\omega_0}=1.0$, $\frac{\hbar\omega_c}{\hbar\omega_0}=0.5$, $\frac{\Gamma}{\Omega}=1.0$, $\frac{\omega_0}{\Omega}=2.0$, $\omega_0\tau_e=1.5$. Also, we use $\hbar\omega_0=1.0$ and $k_B=1.0$}
\end{center}
\end{figure} 
Thus, free energy of the system becomes
\begin{equation}
F(T)=-\frac{2\gamma\Gamma^2\pi^3}{5\Omega^4}\hbar\omega_0^2\Big(\frac{k_BT}{\hbar\omega_0}\Big)^4.
\end{equation}
Entropy is given by
\begin{equation}
S(T)=\frac{8\gamma\Gamma^2\pi^3}{5\Omega^4}k_B\omega_0\Big(\frac{k_BT}{\hbar\omega_0}\Big)^3.
\end{equation}
Thus, the decay behaviour of the entropy is much faster than that of the usual coordinate-coordinate coupling. This is due to the much stronger dependence of the memory friction kernel on the frequency and hence it changes the thermodynamic behaviour of the system a lot at low temperature. Let us discuss, what happens in the absence of the harmonic confining potential. For $\omega_0=0$, $ I_2\stackrel{\omega\rightarrow 0}{\simeq}-\frac{12\gamma(1-\gamma/\Omega)\Gamma^2}{\omega_c^2\Omega^4}\omega^2$, and $I_1=0$. Thus, the free energy can be written down as 
\begin{equation}
F(T)=-\frac{4\gamma(1-\gamma/\Omega)\Gamma^2\pi^3}{45\Omega^4}\hbar\omega_c^2\Big(\frac{k_BT}{\hbar\omega_c}\Big)^4,
\end{equation}
and entropy is given by 
\begin{equation}
S(T)=\frac{16\gamma(1-\gamma/\Omega)\Gamma^2\pi^3}{45\Omega^4}k_B\omega_c\Big(\frac{k_BT}{\hbar\omega_c}\Big)^3.
\end{equation}
\subsection{Radiation Heat Bath}
In this case, the Fourier transform of the associated memory friction function is given by \cite{w} 
\begin{equation}
\tilde{\gamma}_2(\omega)=\frac{2e^2\omega\Omega^{\prime 2}}{3c^3(\omega+i\Omega^{\prime})},
\end{equation}
where $e$ is the charge of the radiation field, $c$ is the velocity of light, $\Omega^{\prime}$ is the large cut-off frequency. Thus,
\begin{eqnarray*}
I_1&=&\frac{3\omega_0^2\tau_e\omega^2+\tau_e^3\omega_0^2\omega^4-\tau_e\omega^4}{\lbrack (\omega_0-\omega^2)^2+\omega^2\omega_0^4\tau_e^2\rbrack(1+\omega^2\tau_e^2)}\\
&&\stackrel{\omega\rightarrow 0}{\simeq}\frac{3\tau_e}{\omega_0^2}\omega^2,
\end{eqnarray*}
where $\tau_e=\frac{2e^2}{3Mc^3}$ and $M=m+\frac{2e^2\Omega^{\prime}}{3c^3}$ = renormalized mass. Similarly one can show that $I_2=0$. Thus, the free energy for the model system with radiation heat bath is given by
\begin{equation}
F(T)=-\frac{\pi^3}{5}\hbar\omega_0^2\tau_e\Big(\frac{k_BT}{\hbar\omega_0}\Big)^4.
\end{equation}
The decay behaviour of the entropy is given by the following expression :
\begin{equation}
S(T)=\frac{4\pi^3}{5}k_B\omega_0\tau_e\Big(\frac{k_BT}{\hbar\omega_0}\Big)^3.
\end{equation}
Comparing equation (60) with equation (65), one can conclude that the decay behaviour of entropy with temperature for the radiation heat bath is same as that of the coordinate (velocity)- velocity (coordinates) coupling scheme. \\
\section{Velocity-Velocity Coupling}
In this section, we consider the second atypical case of dissipation where the velocity of the system is coupled with the velocities of the heat bath. This velocity-velocity (v-v) coupling scheme practically exists in electromagnetic problems such as superconducting quantum interference devices \cite{sch,hanggi} or in electromagnetic field \cite{w}. The spectrum of the friction memory function is completely different from that of coordinate-coordinate coupling and velocity (coordinate)-coordinate (velocities) scheme. The Fourier transform of the memory friction function is given by
\begin{equation}
\tilde{\gamma}_4(\omega)=\frac{2m\gamma\omega^4}{\Gamma^2\omega^2+(\Omega^2-\omega^2)^2}.
\end{equation} 
The expressions for $I_1$ and $I_2$ for this particular scheme are as follows
\begin{eqnarray*}
I_1&=&\frac{2m^2\gamma\omega^4DA+8m^2\gamma\omega^4AC-2m^2\gamma\omega^5A^{\prime}C}{m^2(CA)^2+4m^2\gamma^2\omega^{10}}\\
&&\stackrel{\omega\rightarrow 0}{\simeq}\frac{10\gamma\omega^4}{\omega_0^2\Omega^4}.
\end{eqnarray*}
and
\begin{eqnarray*}
-I_2&=&\frac{2m^2\gamma\omega^4DA+8m^2\gamma\omega^4AC_1-2m^2\gamma\omega^5A^{\prime}C_1}{m^2(C_1A)^2+4m^2\gamma^2\omega^{10}}\\
&&+\frac{2m^2\gamma\omega^4DA+8m^2\gamma\omega^4AC_1-2m^2\gamma\omega^5A^{\prime}C_1}{m^2(C_2A)^2+4m^2\gamma^2\omega^{10}}\\
&&-2\frac{2m^2\gamma\omega^4DA+8m^2\gamma\omega^4AC-2m^2\gamma\omega^5A^{\prime}C}{m^2C^2A^2+4m^2\gamma^2\omega^{10}}\\
&&\stackrel{\omega\rightarrow 0}{\simeq}\frac{10\gamma\omega^4}{\omega_0^2\Omega^4}+\frac{10\gamma\omega^4}{\omega_0^2\Omega^4}-2\frac{10\gamma\omega^4}{\omega_0^2\Omega^4}=0.
\end{eqnarray*}
\begin{figure}[h]
\begin{center}
{\rotatebox{270}{\resizebox{7cm}{12cm}{\includegraphics{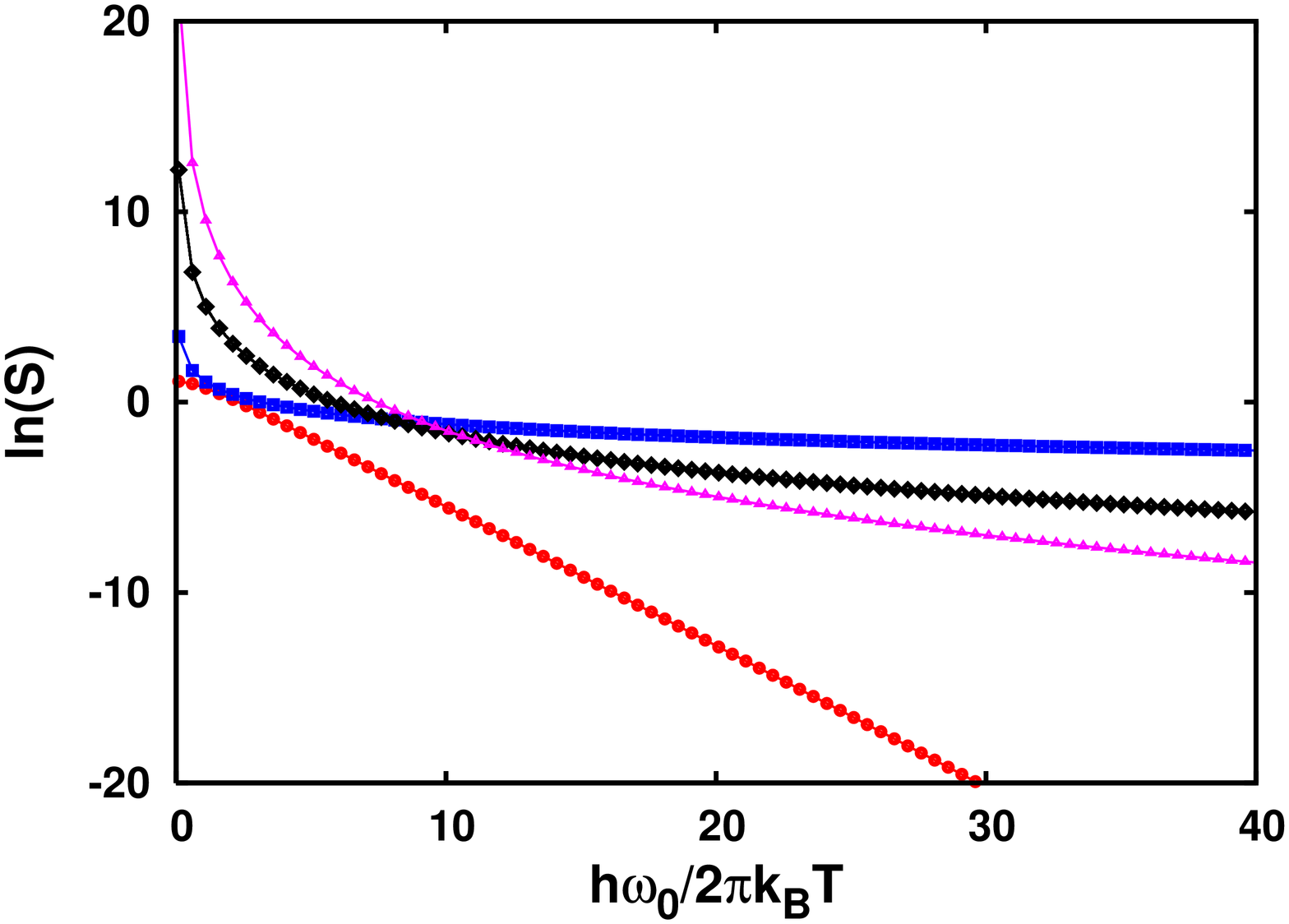}}}}
\caption{(color online) Plot of $\ln(S)$ versus dimensionless inverse-temperature $\frac{h\omega_0}{2\pi k_BT}$ for the without dissipation case (red filled circle), for the coordinate-coordinate (c-c) coupling (blue filled square), for the coordinate-velocity (c-v) coupling (balck filled diamond) and for the velocity-velocity (v-v) coupling (pink filled triangle) schemes. To plot this figure, we use $\frac{\hbar\gamma}{\hbar\omega_0}=1.0$, $\frac{\hbar\omega_c}{\hbar\omega_0}=0.5$, $\frac{\Gamma}{\Omega}=1.0$, $\frac{\omega_0}{\Omega}=2.0$, $\omega_0\tau_e=1.5$. Also we use $\hbar\omega_0=1.0$ and $k_B=1.0$}
\end{center}
\end{figure} 
Thus, the free energy of the charged oscillator in the presence of an external magnetic field for velocity-velocity coupling scheme is given by
\begin{equation}
F(T)=-\frac{144\pi^5}{189}\Big(\frac{\omega_0}{\Omega}\Big)^4\hbar\gamma\Big(\frac{k_BT}{\hbar\omega_0}\Big)^6.
\end{equation}
The decay behaviour of entropy with temperature becomes as follows :
\begin{equation}
S(T)=\frac{864\pi^5}{189\Omega}\Big(\frac{\omega_0}{\Omega}\Big)^3k_B\gamma\Big(\frac{k_BT}{\hbar\omega_0}\Big)^5.
\end{equation} 
Now, we analyze the same case without the harmonic confining potential. For $\omega_0=0$, we have $I_2\stackrel{\omega\rightarrow 0}{\simeq}-\frac{20\gamma(1-\gamma/\Omega)\omega^4}{\omega_c^2\Omega^4}$, and $I_1=0$. Thus, the free energy of the system is given by
\begin{equation}
F(T)=-\frac{96\pi^5}{189}\Big(\frac{\omega_c}{\Omega}\Big)^4\hbar\gamma(1-\gamma/\Omega)\Big(\frac{k_BT}{\hbar\omega_c}\Big)^6,
\end{equation}
and entropy of the system is as follows :
\begin{equation}
S(T)=\frac{576\pi^5}{189\Omega}\Big(\frac{\omega_c}{\Omega}\Big)^3k_B\gamma(1-\gamma/\Omega)\Big(\frac{k_BT}{\hbar\omega_c}\Big)^5.
\end{equation} 
Again as $T\rightarrow0$, entropy vanishes ($S(T)\rightarrow 0$) in conformity with Nernst's theorem. But, the decay behaviour of $S(T)$ is even faster than the coordinate(velocity)-velocity (coordinate) coupling scheme. So, velocity-velocity coupling scheme is the most beneficial coupling scheme to ensure third law of thermodynamics.\\
To demonstrate the distinguishing decay behaviour of free energy for the three kind of coupling schemes, we show the log plot of free energy as a function of the inverse of temperature in figure 2. In figure 3, we show the log-plot of entropy as a function of inverse of temperature. From these plots, one can conclude that the thermodynamical functions of velocity-velocity (v-v) coupling scheme exhibit a markedly faster decaying behaviour than the other two coupling schemes. This can be easily understood by observing their corresponding friction spectra. This is seen that the friction function of the velocity-velocity coupling scheme (Eq. 66) is much more strongly dependent on frequency than the other two schemes. This enables the system to behave much stronger decay behaviour in the low temperature regime.\\
\section{Conclusion}
In this work, we have analyzed the low temperature quantum thermodynamic behaviour in the context of dissipative diamagnetism with different anomalous coupling schemes. The free energy for our system consists of the charged quantum harmonic magneto-oscillator in an arbitrary heat bath is derived by using the ``remarkable" formula of Ford et al \cite{r,s} which involves only a single integral. One can exactly calculate this integral at low temperature limit. Hence, the low temperature thermodynamic functions can be derived explicitly. Mainly, the decay behaviour of entropy with temperature is studied for the charged magneto-oscillator. Thus, the validity of the third law of thermodynamics is established in the quantum regime for the dissipative diamagnetism with atypical coupling.\\
\indent
In the absence of dissipation, thermodynamic functions decay exponentially to zero.
The presence of finite quantum dissipation changes this well known Einstein like behaviour of exponential decay of entropy into a weaker power law dependence in friction and temperature even in the presence of an external magnetic field. Different thermodynamic functions decay much faster with temperature in the presence of anomalous coupling than the usual coordinate-coordinate coupling. It can be concluded from the observation of fast decay of entropy that the velocity dependent coupling is advantageous to ensure third law of thermodynamics. In that sense the velocity-velocity (v-v) coupling is the most helpful scheme to restore third law. It is seen that the thermodynamic entropy for our dissipative diamagnetic system vanishes according to a power law in temperature with the same exponent that characterizes the frequency dependence of the memory friction function in the limit of vanishing frequency ($\omega\rightarrow 0$). For $\omega_0\neq 0$ case, the slope of the decay curve depends on friction, $\gamma$, and the confining harmonic oscillator frequency, $\omega_0$. Also, one can note that low temperature thermodynamic functions are independent of $B$ in all the instances discussed in this work except the case of without dissipation for $\omega\neq 0$. In the absence of confining potential, the decay behaviour of entropy with temperature maintains the same kind of power law as that of $\omega_0\neq 0$. But, the slope of the decay curve for $\omega_0=0$ case depends on $\gamma$, cut-off frequency of the heat bath and on the cyclotron frequency, $\omega_c$. From this analysis we can conclude that quantum dissipation is an integral aspect of nanostructures at very low temperature.\\
\indent
The results obtained from this kind of analysis are not only of theoretical interest but it can be found to be relevant for experiments in nanosciences where one wants to examine the validity of quantum thermodynamics of small systems which are strongly coupled to heat bath \cite{mazenko,datta,imry,chakravarty,x}, in fundamental quantum physics, and in quantum information \cite{bennett,zeilinger,giulini,myatt} .\\  
{
\end{document}